\documentclass[aps,prb,twocolumn,amsmath,amssymb,superscriptaddress,showpacs]{revtex4}
\usepackage{graphics}
\usepackage{graphicx}
\usepackage{bbm}
\usepackage{amsfonts}
\usepackage{amssymb}
\usepackage{epsfig}

\begin{document}

\title{Quantum phase transitions of a generalized compass chain
with staggered Dzyaloshinskii-Moriya interaction}

 \author {Qing-Qiu Wu}

\affiliation{College of Physics, Optoelectronics and Energy, Soochow
University, Suzhou, Jiangsu 215006, P.R. China}

 \author {Wei-Hai Ni}
\affiliation{College of Physics, Optoelectronics and Energy, Soochow
University, Suzhou, Jiangsu 215006, P.R. China}

 \author {Wen-Long You}
 \email{wlyou@suda.edu.cn}
\affiliation{College of Physics, Optoelectronics and Energy, Soochow
University, Suzhou, Jiangsu 215006, P.R. China}

\begin{abstract}
We consider a class of one-dimensional compass models with staggered Dzyaloshinskii-Moriya exchange interactions in an external transverse magnetic field. Based on the exact solution derived from Jordan-Wigner
approach, we study the excitation gap, energy spectra, spin correlations and critical properties at phase transitions. We explore mutual effects of the staggered Dzyaloshinskii-Moriya interaction and the magnetic field on the energy spectra and the ground-state phase diagram. Thermodynamic
quantities including the entropy and the specific heat are discussed, and their universal scalings at low temperature are demonstrated.
\end{abstract}


\maketitle

\section{Introduction}
 As an old yet flourishing concept \cite{Dzyaloshinskii58,Moriya60}, the Dzyaloshinskii-Moriya (DM) interaction has received immense attentions recently.  Such inversion-symmetry-breaking exchange is believed to be an indispensable ingredient in understanding spin glasses\cite{Fert80}, the magnetism-driven ferroelectricity in multiferroics \cite{Cheong07,Hosho05,Tokura10}, the chiral texture \cite{Heinze11,Luchaire16,Boulle16,Jiang15}. 
The DM antisymmetric interaction
is an indirect exchange interaction between two neighboring magnetic cations, which is transferred via anion neighbors by spin-polarized conduction electrons with a spin-orbit coupling. Usually it emerges
in those compounds with either low or distorted crystal symmetries, and recently
major attention has been paid to the nature of the DM interaction at transition-metal (TM) interfaces \cite{Bode07}. The interfacial
DM interaction gives rise to several exotic magnetic phases.

In light of most analyses focusing on uniform fields,
DM interactions can also be nonuniform in realistic situation, especially in low dimensions. A magnetic chain with a low symmetry of the crystal structure will inevitably have twisting arrangements of atoms. The different orientations of successive ligands essentially lead to a zigzag alignment of adjacent ligands. In this connection there is
increasing interest in the effects of the staggered field motivated by the experimental work on a number of
materials.  For instance, in the Ising-like antiferromagnetic compound CsCoCl$_3$ and CsCoBr$_3$ \cite{Goff95}, the magnetic cations are surrounded by trigonally distorted octahedra of halogen anions. The triangular distortion of the crystal field renders two crystallographically inequivalent magnetic ions along the chain with a zigzag orientation. 
 This collectively mediates an effective staggered DM field upon the background of a uniform field. In contrast to the uniform DM interaction, which tends to provoke a gapless chiral phase under a saturation uniform DM field \cite{Jafari08,You2,Liu15}, the staggered interactions are inclined to induce a dimerized gap. The staggered DM interaction in Cu benzoate antiferromagnetic chain plays important role in understanding gap formation\cite{Affleck99,Oshikawa97,Zhao03}. Note that the sign of the DM interaction 
 can be determined by using synchrotron radiation \cite{Dmitrienko14}.

Meanwhile, the progress in the experimental front is achieved by synthesis of magnetic quasi-one dimensional systems \cite{Schauss15}. One-dimensional (1D) quantum models are natural playgrounds for studying various phases and some novel magnetic properties, especially some exactly solvable models. A representative model is compass model, in which Ising-like interactions on different spatially bonds prefer nonparallel easy axes. In such a highly frustrated quantum
system, the spins cannot order simultaneously to minimize all local
interactions.  

The outline of the paper is as follows: in Section \ref{sec:Ham}
we present the Hamiltonian and its analytical solution through Jordan-Wigner approach. In Section \ref{sec:QPT} we
investigate mutual effects of a staggered DM interaction and an external magnetic field on the ground state properties of the model. Quantum phase transitions (QPTs) are identified through energy spectra and spin correlations. Then the entropy and specific heat are supplemented in Section \ref{sec:Thermodynamics}. Finally we conclude and summarize our results in Section \ref{sec:Con}.

\section{Hamiltonian and Solution}
\label{sec:Ham}
A 1D generalized compass model (GCM) was a microscopic model to
mimic zigzag spin chains in perovskite TM oxide. 
The model is given by \cite{You1,You2,You16}
\begin{eqnarray}
\cal{H}_{\rm GCM}&=& \sum_{i=1}^{N^\prime}\left\{
J_{o}\tilde{\sigma}_{2i-1}(\theta)\tilde{\sigma}_{2i}(\theta)
+J_{e}\tilde{\sigma}_{2i}(-\theta)\tilde{\sigma}_{2i+1}(-\theta)\right\},\nonumber \\
\label{Hamiltonian1}
\end{eqnarray}
where the operator with a tilde sign is defined as linear combinations of
$\{\sigma_{l}^x,\sigma_{l}^y\}$ pseudospin components,
\begin{eqnarray}
\tilde{\sigma}_{i}(\theta)&\equiv& \cos(\theta/2)\,\sigma_{i}^x
+\sin(\theta/2)\,\sigma_{i}^y.
\end{eqnarray}
Here, $N'=N/2$ is the number of two-site unit cells. $J_o$ ($J_e$) denotes the amplitude of the nearest-neighbor intrachain interaction on odd (even)
bonds. A variable $\theta$ is introduced to specify the angle difference between the easy axes on odd and even bonds, like a 31$^\circ$ canting angle of Co$^{2+}$ ions from the $c$ axis of CoNb$_2$O$_6$ system \cite{Mochizuki11}.

The DM interaction in our paper takes the following form:
\begin{eqnarray}
{\cal H}_\mathrm{DM}=\sum_{i=1}^{N} \vec{E}_{i,i+1} \cdot (\vec{\sigma}_i \times \vec{\sigma}_{i+1}). \label{Hamiltonian_DM}
\end{eqnarray}
In Eq.(\ref{Hamiltonian_DM}) a possibly spatially inhomogeneous DM interaction vector $\vec{E}_{i,i+1}$ is associated with the bond between magnetic moments $\vec{\sigma}_i$ and $\vec{\sigma}_{i+1}$. Such interactions emerge in the low-dimensional magnetic system lacking structural inversion symmetry, e.g., near magnetic surfaces/edges, or they can appear as an energy current of 1D compass chain in the presence of the magnetic field \cite{Qiu16}.  Consequently,
a staggered DM interaction is generated by the energy current induced from the magnetic field, i.e.,
$\vec{E}_{2i-1,2i}$= $E J_o \hat{z}$, and $\vec{E}_{2i,2i+1}$= $E J_e \hat{z}$.
Further, an external uniform magnetic field is incorporated,
\begin{eqnarray}
{\cal H}_{\rm h}= h\hat{z}  \cdot \sum_{i=1}^{N}   \vec{\sigma}_{i}.
\end{eqnarray}

The competition among these complex interactions essentially enriches the ground-state phase diagram of GCM. For example, the mutual effect of the DM interaction and the transverse magnetic field produce an incommensurate chiral state in CsCuCl$_3$ \cite{Schotte98}. To this end, putting the exchange interaction and the staggered DM interaction together with the magnetic field promises to deliver something interesting:
\begin{eqnarray}
{\cal H}&=& \cal{H}_{\rm GCM} + {\cal H}_{\rm h}+ {\cal H}_\mathrm{DM}.
\label{model}
\end{eqnarray}

The Jordan-Wigner transformation
maps explicitly between quasispin operators and spinless fermion
operators \cite{EBarouch70}:
\begin{eqnarray}
\sigma_j^z=1-2c_j^\dagger c_j,  \sigma_j^x=e^{i\phi_j} (c_j^\dagger + c_j), \sigma_j^y=i e^{i\phi_j} (c_j^\dagger - c_j),
\end{eqnarray}
with $\phi_j$ being the phase string defined as $\phi_j=\pi \sum_{l<j}c_l^\dagger c_l$.
 Consequently, we have a simple bilinear
form of Hamiltonian in terms of spinless fermions:
\begin{eqnarray}
{\cal H}&=&
 \sum_{i=1}^{N'} \left[J_{o}  e^{i\theta} c_{2i-1}^{\dagger} c_{2i}^{\dagger}
  + J_{o}(1-2 i E ) c_{2i-1}^{\dagger} c_{2i} \right. \nonumber \\
&+& J_{e}  e^{-i\theta} c_{2i}^{\dagger} c_{2i+1}^{\dagger}
+ J_{e}(1-2 i E) c_{2i}^{\dagger} c_{2i+1} +{\rm H.c.}  \nonumber \\
&+& \left.  h(1-2 c_{2i-1}^{\dagger} c_{2i-1})+h(1-2 c_{2i}^{\dagger} c_{2i})  \right].
\label{Hamiltonian2}
\end{eqnarray}
The dual fermionic model describes a $p$-wave
superconductor, and sometimes also was dubbed as an extended Su-Schrieffer-Heeger model, which has raised lots of research interests \cite{Tong15,Jafari17}.

And next a discrete Fourier transformation for odd/even spin sites is
applied in the following way,
\begin{eqnarray}
c_{2j-1}=\frac{1}{\sqrt{N'}}\sum_{k}e^{-ik j}a_{k},\text{ \ \ }c_{2j}=%
\frac{1}{\sqrt{N'}}\sum_{k}e^{-ik j}b_{k},
\end{eqnarray}
with the discrete momenta given as follows,
\begin{eqnarray}
k=\frac{n\pi}{ N^\prime  }, \quad n= -(N^\prime\!-1), -(N^\prime\!-3),
\ldots, (N^\prime\! -1). \quad
\end{eqnarray}
The Hamiltonian takes the following form which is suitable to
introduce the Bogoliubov transformation,
\begin{eqnarray}
\hat{H}_{E}&=&\! \sum_{k} \left[ A_k a_{k}^{\dagger} b_{k}
+\! B_k a_{k}^{\dagger}b_{-k}^{\dagger} -\! A_k^* a_{k}b_{k}^{\dagger}
-\!B_k^* a_{k}b_{-k} \right.\nonumber \\
&+&
\! \left. h (a_{-k} a_{-k}^{\dagger}- a_{k}^{\dagger} a_{k}) + h (b_{-k} b_{-k}^{\dagger}- b_{k}^{\dagger} b_{k})  \right].\quad \quad
\label{Hamiltonian5}
\end{eqnarray}
Here
\begin{eqnarray}
A_k&=& J_{o}(1-2i E)+ J_{e}(1+ 2i E) e^{ik},  \nonumber \\
B_k&=& J_oe^{i\theta}-J_e e^{i(k-\theta)}.
\end{eqnarray}
To diagonalize the Hamiltonian Eq. (\ref{Hamiltonian5}), we rewrite
it in the Bogoliubov-de Gennes (BdG) form in terms of Nambu spinors:
\begin{eqnarray}
\cal{H} &=&  \sum_{k}
\Gamma_k^{\dagger}
\hat{M}_k
\Gamma_k, \label{FT2}
\end{eqnarray}
where
\begin{eqnarray}
\hat{M}_k
&=&\frac{1}{2} \left(\begin{array}{cccc}
 -2 h &  A_k &  0   &  B_k  \\
A_k^* &  -2 h & -B_{-k}    & 0     \\
0 &  -B_{-k}^*    & 2 h & -A_{-k}^*\\
 B_k^*   & 0   &  -A_{-k} & 2h
\end{array}\right),\label{Mk}
\end{eqnarray}
and $\Gamma_k^{\dagger} =(a_{k}^{\dagger},b_k^{\dagger},a_{-k},b_{-k})$. 

\section{Quantum Phase transition}
\label{sec:QPT}
After diagonalizing Hamiltonian Eq.(\ref{Mk}),
we obtain four branches of energy spectra $\varepsilon_{k,j}$, ($j$=1, $\cdots$,
4). A unitary transformation $\hat{U}_k$ can transform the Hermitian matrix
(\ref{Mk}) into a diagonal form,
\begin{eqnarray}
\hat{\Upsilon}_k=\hat{U}_k \hat{M}_k \hat{U}_k^{\dagger}.
\end{eqnarray}
The quasiparticle (QP) operators,
$\{\gamma_{k,1}^{\dagger},\gamma_{k,2}^{\dagger},\gamma_{k,3}^{\dagger},
\gamma_{k,4}^{\dagger}\}$, are connected with
$\{a_k^{\dagger},a_{-k}^{},b_k^{\dagger},b_{-k}^{}\}$ through the
following relation,
\begin{eqnarray}
\left(
\begin{array}{c}
\gamma_{k,1}^{\dagger} \\
\gamma_{k,4}^{\dagger}  \\
\gamma_{k,2}^{ \dagger}\\
\gamma_{k,3}^{\dagger}
\end{array}
\right)=\hat{U}_{k} \left(
\begin{array}{c}
a_k^{\dagger}  \\
a_{-k}   \\
b_k^{\dagger}   \\
b_{-k}
\end{array}%
\right). \label{eq:2DXXZ_RDM}
\end{eqnarray}%

\begin{figure}
 \includegraphics[width=9cm]{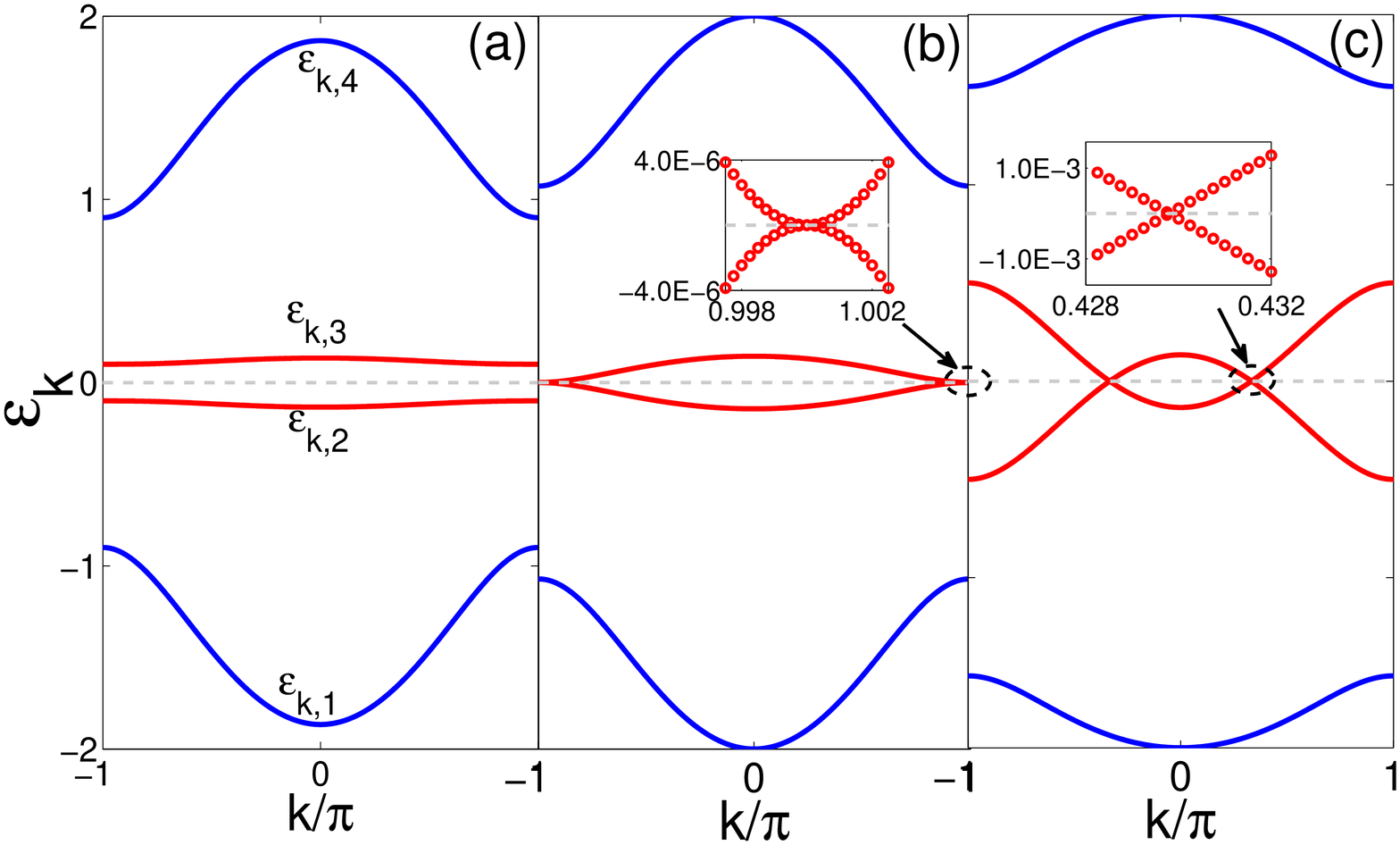}
 \caption{(Color online) The energy spectra $\varepsilon_{k,j}$, ($j$=1, $\cdots$,4) for for increasing $E$: (a) $E=0.2$; (b)$E=0.25$ and (c) $E=0.4$. The insets in (b)-(c) are the amplifications of the corresponding dashed circles. Parameters are as follows: $J_o$=1, $J_e$=1, $\theta=\pi/3$, $h$=0.  }
  \label{Spe0}
 \end{figure}

\begin{figure}
 \includegraphics[width=9cm]{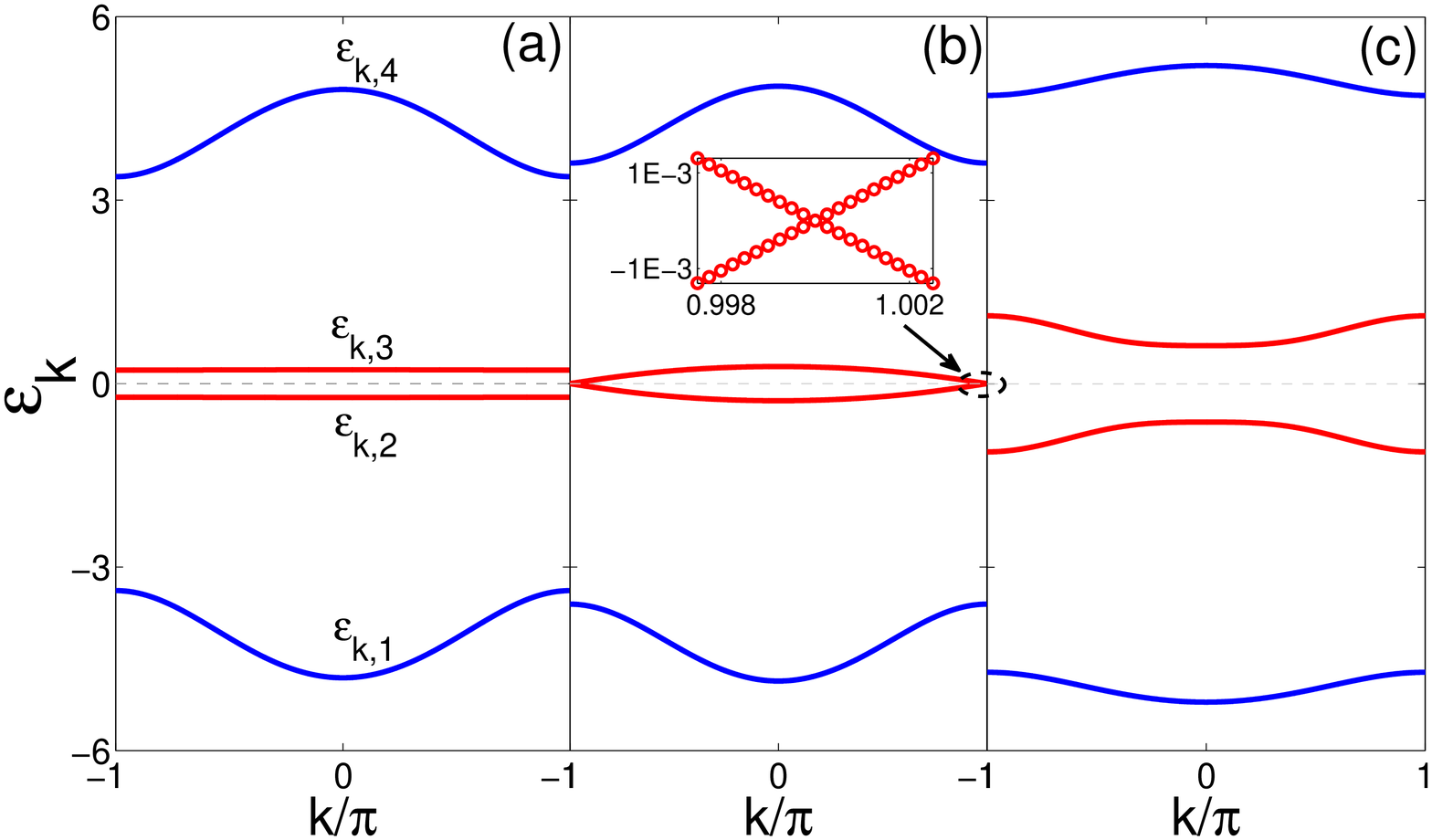}
 \caption{(Color online) The energy spectra $\varepsilon_{k,j}$, ($j$=1, $\cdots$,4) for for increasing $E$: (a) $E=0.1$; (b)$E=0.2$ and (c) $E=0.5$. The inset in (b) is the amplification of the dashed circle
below. Parameters are as follows: $J_o$=1, $J_e$=4, $\theta=\pi/3$, $h$=0.  }
  \label{Spe1}
 \end{figure}

A zero field $h$=0, the analytical solutions can be retrieved:
\begin{eqnarray}
\varepsilon_{k,1(2)}=
-\frac{1}{2}\sqrt{\varsigma_k\pm \sqrt{\varsigma_k^2-\tau_k^2}},\label{e1} \\
\varepsilon_{k,3(4)}=
 \frac{1}{2}\sqrt{\varsigma_k\mp \sqrt{\varsigma_k^2-\tau_k^2}},
\label{e2}
\end{eqnarray}
where
\begin{eqnarray}
\varsigma_k&=&\frac{1}{2}\left(\vert A_k \vert^2 + \vert A_{-k} \vert^2 + \vert B_k
\vert^2+ \vert B_{-k} \vert^2\right), \nonumber \\
\tau_k&=&\vert  A_k A_{-k}^* - B_k B_{-k}^* \vert.
\end{eqnarray}
From Eqs.(\ref{e1}-\ref{e2}) one finds the spectra are
not invariant with respect to the $k\to -k$ transformation and also energy $\varepsilon=0$,  as is evidenced in Figs.
\ref{Spe0} and \ref{Spe1}.  The most important properties of the 1D
 spin system are manifested in the ground state. Accordingly, the bands with positive energies correspond to the electron excitations, while the negative ones are the corresponding
hole excitations and then occupied.  The ground-state
 energy density of our model can be written as
 \begin{eqnarray}
e_0 &=& \frac{1}{N'}\sum_{k} (\varepsilon_{k,1} +\varepsilon_{k,2})= -\frac{1}{\sqrt{2}N'} \sum_{k} \sqrt{\varsigma_k + \tau_k }. \quad
 \label{E0}
 \end{eqnarray}
The spectral gap is determined by the absolute value of the
difference between the second and the third energy branch,
\begin{equation}
\Delta=\min_{k}\vert  \varepsilon_{k,2}- \varepsilon_{-k,3}\vert.
\end{equation}
The gap closes at some critical momentum $k_c$ delimited by $\tau_{k_c}$=0. The critical mode occurs at $k_c= \pm \pi$ and the phase boundary is given by
\begin{equation}
 E_c  =  \pm  \sqrt{J_o J_e}\cos \theta/(J_o+J_e) .
 \end{equation}
$\varepsilon_{k,3}$ and $\varepsilon_{k,2}$ touch $\varepsilon$=0 at modes $k=\pm \pi$ when $E=E_c$.
The critical behaviour is determined by those low-energy states near the critical modes. The scaling of energy scales to length scales, i.e., $\varepsilon_{k}\sim L^{-z}$, define a dynamic critical exponent $z$. As show in insets of Fig.\ref{Spe0}(b) and Fig.\ref{Spe1}(b), the spectra vanish quadratically at $\pm \pi$ corresponding to a dynamical exponent
$z = 2$. For $J_o$$\neq$$J_e$, the gap opens again by continuously increasing $E$ [see Fig.\ref{Spe1}(c)]. However, a further enhancement of $E$ makes the system remain gapless for $J_o$=$J_e$ as a consequence of bands inversion; $\varepsilon_{k,3}$ and $\varepsilon_{k,2}$ cross at two generally incommensurate and symmetric
momenta $\pm k_{ic}$, which are given by
\begin{equation}
k_{\rm ic}=\arccos\left[\frac{2 J_o J_e E^2 -J_o J_e \cos^2 \theta}{ E^2 (J_o^2+J_e^2)}\right].
\end{equation}
In this case, the spectra around critical modes $k_{\rm c}$ are relativistic, implying a dynamical exponent $z=1$ in the Tomonaga-Luttinger-liquid phase [see Fig.\ref{Spe0}(c)].

\begin{figure}[t!]
\includegraphics[width=8cm]{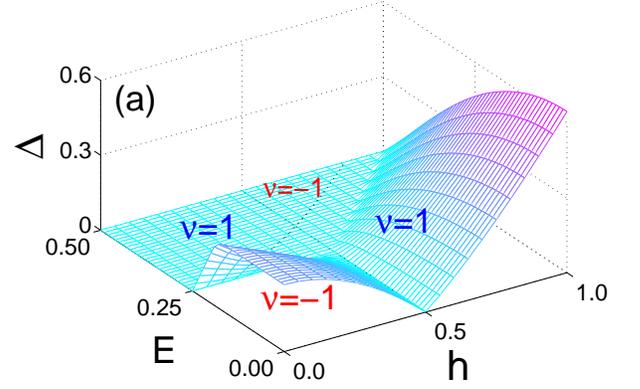}
\includegraphics[width=8cm]{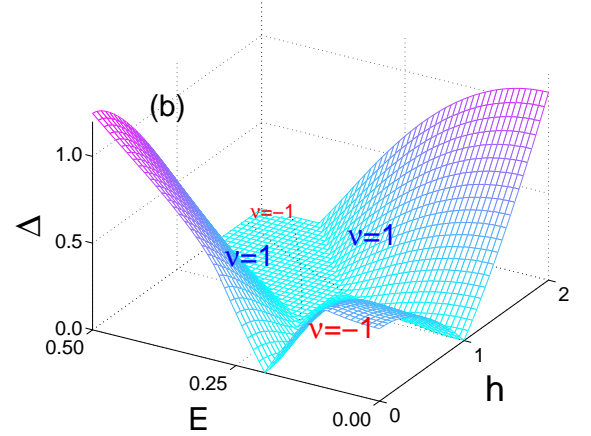}
\caption{(Color online)
The gap $\Delta$ as a function of $E$ and $h$ for (a) $J_e$=1 and (b) $J_e$=4.
Parameters are as follows: $J_o$=1, $\theta$=$\pi/3$. The dotted lines delimitate the topological protected number $\nu$. }
\label{Gap2}
\end{figure}

In the presence of a finite magnetic field, which breaks the time-reversal symmetry, the analytical expressions of eigenspectra $\varepsilon_{k,j}$
($j=1,\cdots, 4$) are rather lengthy and cannot be given in an explicit form.
However, Hamiltonian Eq. (\ref{Mk}) still respects an artificial particle-hole symmetry. This system
belongs to topological class $D$ with topological invariant $Z_2$
in one dimension \cite{Altland,Chiu16}, which satisfies ${\cal C}^{-1} \hat{M}(-k) {\cal C}=- \hat{M}(k)$. Here particle-hole operator $ {\cal C} = \tau^x \otimes \sigma^0\cal{K}$, where
$\tau^x$ and $\sigma^0$ are the Pauli matrices
acting on particle-hole space and spin space,
respectively, and $\cal{K}$ is the complex conjugate operator.
To be specific, $\varepsilon_{k,4}$=-$\varepsilon_{-k,1}$,
$\varepsilon_{k,3}$=-$\varepsilon_{-k,2}$. Simultaneously $\gamma_{k,4}^{\dagger}$=$\gamma_{-k,1}$,
$\gamma_{k,3}^{\dagger}$=$\gamma_{-k,2}$.
Along these lines,
 we obtain the diagonal form of the Hamiltonian from Eq.(\ref{Mk}),
\begin{eqnarray}
{\cal H}=\sum_{k}\sum_{j=1}^{4}   \varepsilon_{k,j}
\gamma_{k,j}^{\dagger}\gamma_{k,j}^{} .
\label{diagonalform}
\end{eqnarray}
When all quasiparticles above the Fermi surface are
absent the ground-state energy density for the particle-hole excitation spectrum may be expressed as:
\begin{eqnarray}
e_0 = - \frac{1}{N}\sum_{k} \sum_{j=1}^4 \vert \varepsilon_{k,j} \vert.
\label{E0expression}
\end{eqnarray}
The gap for $\theta=\pi/3$ is plotted as a function of $E$ and $h$ in Fig.\ref{Gap2}. One can see there are 3 phases for $J_o $=$J_e$ and 4 phases for $J_o $ $\neq$ $J_e$. In the latter case, a dimerized gap is formed when $E$ is above a critical value.

In terms of particle-hole operator ${\cal C}$, an auxiliary function $W(k)= \hat{M}_{4\times 4}(k) {\cal C} $ is incorporated with skew symmetric form $W(k)^T = -W(-k)$. 
The topological nature of the ground state can be characterized by the Pfaffian of the Hamiltonian at particle-hole symmetric momenta $k=0$ and $\pi$, with $\nu=sgn(Pf(W(0)) Pf(W(\pi)))$, in which $\nu$ is a topological protected number. 
An insulating phase with $\nu$ = -1
corresponds to a topological nontrivial phase\cite{Kitaev01,Ghosh10}, and it is a topological trivial phase otherwise. The Pfaffian can be obtained straightforwardly: $ Pf[W(0)] $=$   E^2(J_e - J_o)^2 -h^2 + J_e J_o \cos^2 \theta   $ and $ Pf[W(\pi)] $=$  E^2(J_e + J_o)^2 - h^2 - J_e J_o \cos^2 \theta $. The values of $\nu$ are incorporated in Fig.\ref{Gap2}. One can see a topological nontrivial phase exists for small $E$ and $h$.

In order to characterize the QPTs, we
study nearest neighboring spin correlation functions $C_e^{\alpha}$ ($C_o^{\alpha}$)
on even (odd)  bonds  defined by
\begin{eqnarray}
C^{\alpha}_{o(e)}&=&-\frac{2}{N}\sum_{i=1}^{N/2}\langle
\sigma_{2i-1(2i)}^\alpha
\sigma_{2i(2i+1)}^\alpha \rangle,
\end{eqnarray}
where the superscript $\alpha=x,y,z$ denotes the cartesian component,
and chirality correlation function,
\begin{eqnarray}
X_{o(e)}^{\alpha}&=&-\frac{2}{N}\sum_{i=1}^{N/2}\langle
\vec{\alpha} \cdot
(\vec{\sigma}_{2i-1(2i)}\times \vec{\sigma}_{2i(2i+1)} )\rangle, \label{chir}
\label{chire}
\end{eqnarray}
where $\vec{\alpha}$ denotes the unit vector in the direction of a
cartesian component $\alpha$.
The chirality $X^{\alpha}$
(\ref{chir}) will exhibit a sign change under the parity operation but
stay invariant under the time-reversal operation. Within Katsura-Nagaosa-Balatsky (KNB) mechanism, the DM interaction was considered as an essential role in the ferroelectricity \cite{Hosho05,Tokura10}. The electric polarization $\vec{P}$ is
generated by the displacement of oppositely
charged ions in the following way,
\begin{eqnarray}
\vec{P}_{i} \propto \gamma \hat{e}_{ij} \times (\vec{\sigma}_i
\times \vec{\sigma}_j),
\label{polarization}
\end{eqnarray}
where $\hat{e}_{ij}$ is the unit vector connecting the neighboring
spins $\vec{\sigma}_i$ and $\vec{\sigma}_j$, and the coupling coefficient
$\gamma$ of the cycloidal component is material-dependent
\cite{Sergienko06}. One can find $X^{z}$ is equivalent to $y$-component polarization $P^y$ considering $\hat{e}_{ij}=\hat{x}$ .

\begin{figure}[t!]
\includegraphics[width=9cm]{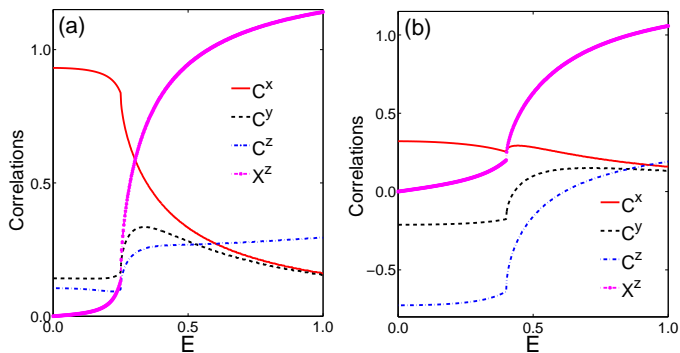}
\caption{(Color online) Evolution of the nearest-neighbor correlations $C^\alpha$ and chirality
$X^\alpha$ for increasing
DM field for
(a) $h=0.2$;
(b) $h=0.8$.
Parameters are as follows: $J_o=1$, $J_e=1$,
$\theta=\pi/3$. }
\label{Orderparameters-Je=1}
\end{figure}

\begin{figure}[t!]
\includegraphics[width=9cm]{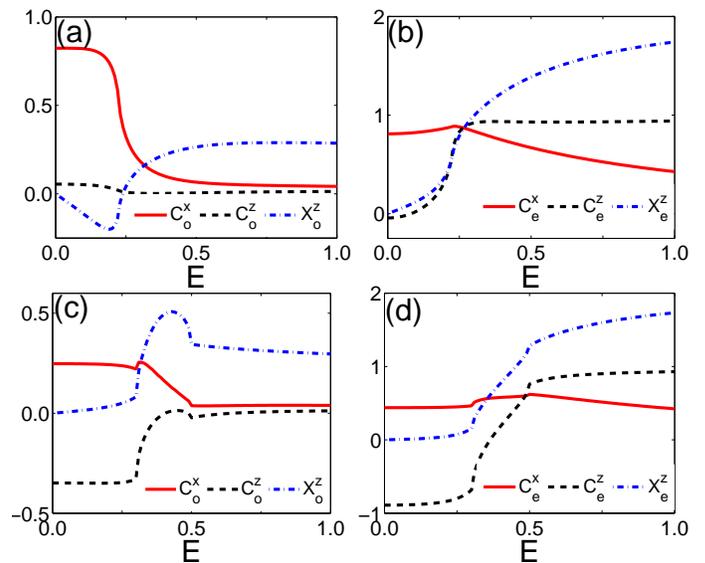}
\caption{(Color online) Evolution of the nearest-neighbor correlations $C^\alpha$ and chirality
$X^\alpha$ for increasing
DM field on (a) odd bonds and (b)even bonds
with $h=0.5$; on (c) odd bonds and (d)even bonds
with $h=1.5$.  Parameters are as follows: $J_o=1$, $J_e=4$,
$\theta=\pi/3$. }
\label{Orderparameters-Je=4}
\end{figure}

The correlation functions for a few typical paths are exhibited in Fig. \ref{Orderparameters-Je=1} for $J_o =J_e$ and Fig. \ref{Orderparameters-Je=4} for $J_o$ $\neq$ $J_e$. In the absence of the magnetic field
($h$=0) and the DM interaction ($E$=0),  the ground state has long range order, which is characterized by the nearest neighbor
correlation functions, among which $x$-components $\{C_l^x\}$ dominate in Fig.\ref{Orderparameters-Je=1}(a) and Fig.\ref{Orderparameters-Je=4}(a-b). This suggests the adjacent spins are antiparallel with a canted angle with respect to the $x$ axis. In a word, the system is in a canted N\'eel (CN) phase.
The external magnetic field $h$ will induce a spin-flop
transition and polarize spins orienting along the $z$ direction \cite{You1}. Such polarized phase is characterized by negative $C_o^{z}$ and $C_e^{z}$, as shown in Fig.\ref{Orderparameters-Je=1}(b) and Fig.\ref{Orderparameters-Je=4}(c-d).

With increase of $E$, different behaviors take place under uniform and staggered DM interactions. For $J_o$ $\neq$  $J_e$, $M^{z}$,  $C_o^{x}$,
$C_o^{y}$, $C_o^{z}$ decrease quickly. On the contrary, $C_e^{x}$,
$C_e^{y}$, $C_e^{z}$ are enhanced. $C_e^{x}$,
$C_e^{y}$ turn to decline after $E$ crosses a threshold value and the system is in a gapped chiral phase, in which the $z$-component chirality
$X^{z}$ starts to grow and dominates over other correlations, as is disclosed in Fig. \ref{Orderparameters-Je=4}(a-b). $C_o^{z}$ and $C_e^{z}$ unexpectedly become saturated in such a phase. This implies that the DM interaction induces spins
to be cycloidally oriented in the $(\sigma^x,\sigma^y)$ easy plane, but spins on the strong bonds especially develop $z$-component antiferromagnetic correlations. The competition of $E$ and $h$ will lead to a Tomonaga-Luttinger-liquid phase, in which
$C_o^{z}$ and $C_e^{z}$ undergo a sign change, as shown in Fig. \ref{Orderparameters-Je=4}(c-d). For $J_o$=$J_e$, the gapped chiral phase does not exist. The ground state will transit into a gapless chiral phase as long as $E$ surpasses a critical value, and the chiralities  $X_o^{z}$ and $X_e^{z}$ grow fleetly.

\begin{figure}[t!]
\includegraphics[width=9cm]{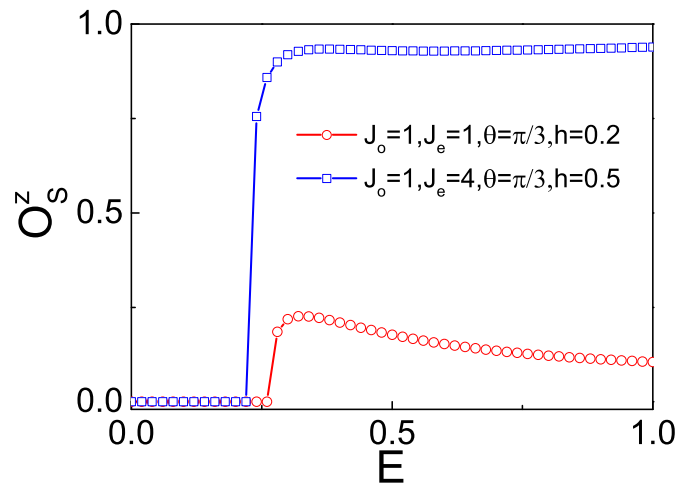}
\caption{(Color online) The string order parameters $O^{z}_s$ versus $E$.
Here $n - k  = 100$ and the bond dimension is set as $\chi=30$. }
\label{string}
\end{figure}

To further analyze the nature of chiral phase, we calculate the $z$-th component string order parameter:
\begin{eqnarray}
 O_{s}^{\alpha} = \langle
\sigma^{\alpha}_{2k} \sigma^{\alpha}_{2k+1}\sigma^{\alpha}_{2k+2}\sigma^{\alpha}_{2k+3}
\cdots \sigma^{\alpha}_{2n}
\sigma^{\alpha}_{2n+1}  \rangle.
\end{eqnarray}
We adopt an infinite time-evolving block decimation (iTEBD) algorithm
\cite{Vidal07,Liu13}, which allows one to solve for the ground state
properties of a 1D translationally invariant spin system of infinite
length. A crucial quantity during this strategy is the
bond dimension $\chi$, i.e., the cut-off dimension of Schmidt
coefficients during singular value decomposition process. Figure  \ref{string}  reveals that a nonlocal correlation $O_s^z$ arises in the both gapless and gapped chiral phases. The presence
of a nonzero string order parameter indicates a
hidden symmetry breaking through the Kramers-Wannier dual transformation \cite{Liu13}, which can be related to a
symmetry-protected topological order.

\section{Thermodynamics }
\label{sec:Thermodynamics}
So far our study focuses on the ground state.
In practice, we can only work at low but finite temperature,
as close to absolute zero as possible, and the finite temperature properties is important theoretically and experimentally.
Thanks to the exact solution of the GCM, it is straightforward
to obtain its full thermodynamic properties at finite temperature.  For the particle-hole excitation spectrum (\ref{diagonalform}),
the free energy of the quantum spin chain at temperature $T$ reads (here and below we use the units with the Boltzmann constant $k_{\rm B}\equiv 1$),
\begin{eqnarray}
{\cal F}= -    T \sum_k\sum_{j=1}^4
\ln\left(2\cosh\frac{\varepsilon_{k,j}}{2  T}\right). \label{freeenergy}
\end{eqnarray}

\begin{figure}
\includegraphics[width=9cm]{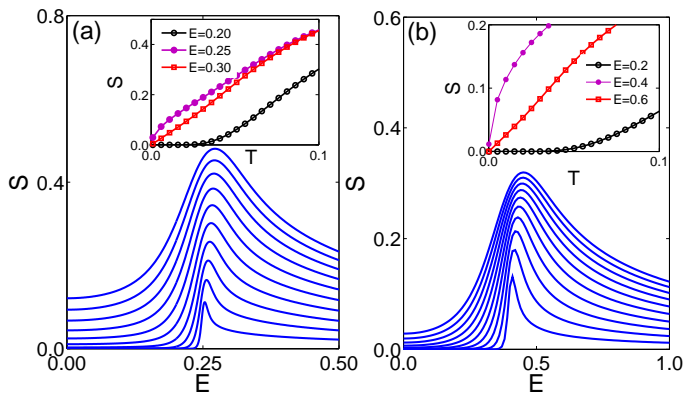}
 \caption{(Color online)
The entropy ${\cal S}$ versus DM field at increasing temperature
$T=0.01,0.02,\cdots,0.1$ (from bottom to top)  for (a) $h$=0 and (b) $h$=0.8. Inset shows the scaling of entropy with respect to $T$ for typical values of $E$. Parameters are as follows: $J_o$=1, $J_e$=1, $\theta=\pi/3$.}
 \label{Fig:Sent1}
\end{figure}

\begin{figure}
\includegraphics[width=9cm]{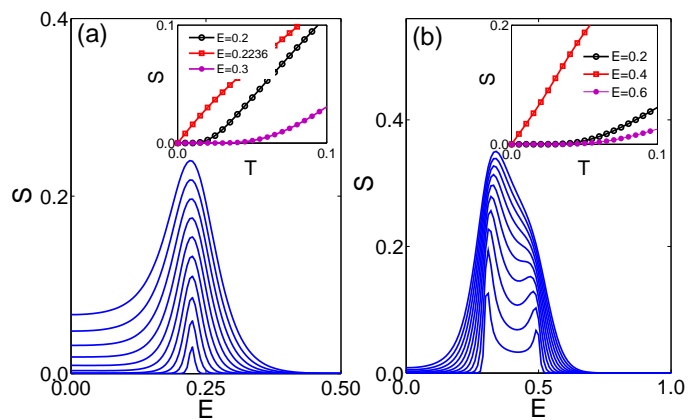}
 \caption{(Color online)
The entropy ${\cal S}$ versus DM field at increasing temperature
$T=0.01,0.02,\cdots,0.1$ (from bottom to top)  for (a) $h$=0.5 and (b) $h$=1.5. Inset shows the scaling of entropy with respect to $T$ for typical values of $E$. Parameters are as follows: $J_o$=1, $J_e$=4, $\theta=\pi/3$.}
 \label{Fig:Sent2}
\end{figure}

Among many
thermodynamic quantities, the entropy ${\cal S}$ provides fundamental information about the evolution of spectra
with increasing the temperature $T$. The complete entropic landscape was recently quantitatively measured
for Sr$_3$Ru$_2$O$_7$ under magnetic field in the vicinity of quantum
criticality \cite{Rost09}. It can be derived from
the free energy ${\cal F}$ (\ref{freeenergy}) via the following thermodynamic relation,
\begin{eqnarray}
{\cal S}&\!=&-\left(\frac{\partial{\cal F}}{\partial T}\right)
\nonumber  \\
&\!=&\!\sum_k\sum_{j=1}^{4}\ln\left(2\cosh\frac{\varepsilon_{k,j}}{2T}\right)
-\sum_k\sum_{j=1}^{4}\left( \frac{\varepsilon_{k,j}}{2T}
\tanh\frac{\varepsilon_{k,j}}{2T}\right). \nonumber \\
\end{eqnarray}
Figure \ref{Fig:Sent1} demonstrates the entropy ${\cal S}$ as a function of the DM field $E$ and the temperature $T$ when the DM interaction is uniform. The entropy  ${\cal S}$ is vanishing asymptotically at zero temperature, and it grows with increasing $T$ when thermal excitations gradually
include more and more excited states. One finds the entropy takes on a distinct maximum for increasing $E$ across a quantum critical point (QCP) at low temperature, annotating that the system is maximally undecided to select a state among the competing phases. The maximum get broader by increasing the temperature.
The entropy shows an exponential activation with $T$ in the gaped phases, i.e., ${\cal S}\propto\exp(-\Delta/T)$, as is indicated in insets of Fig.\ref{Fig:Sent1}.  A scrutiny reveals that the entropy
presents a linear dependence on $T$, i.e., ${\cal S}\propto T$, for low temperatures in gapless phase, while ${\cal S}\propto \sqrt{T}$ at QCPs. This establishes a relation ${\cal S}$ $\propto T^{d/z}$ in the scaling theory to quantum criticality (here the spatial dimension $d$ is 1). This power-law
dependence arises directly from the density of states in one dimension, $D(\epsilon)\propto \epsilon^{d/z-1}$.
For the staggered DM interaction, the entropy ${\cal S}$ in Fig.\ref{Fig:Sent2} also exhibits local maxima across the QCPs. The entropy ${\cal S}(T)$ as a function of temperature $T$ also follows an identical scaling as Fig.\ref{Fig:Sent1} as long as the Fermi-surface topology is guaranteed.


\begin{figure}[t!]
\includegraphics[width=8cm]{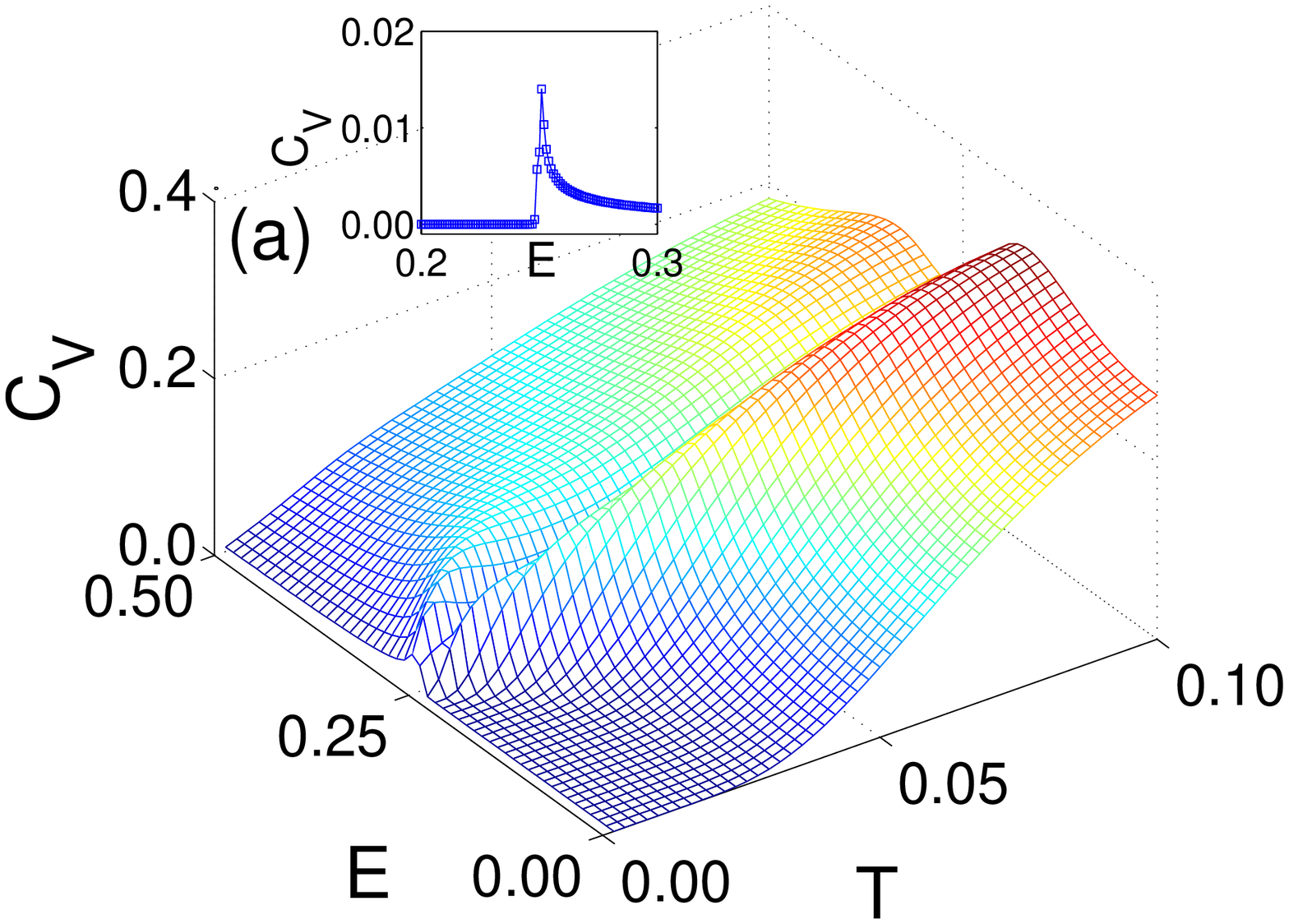}
\includegraphics[width=8cm]{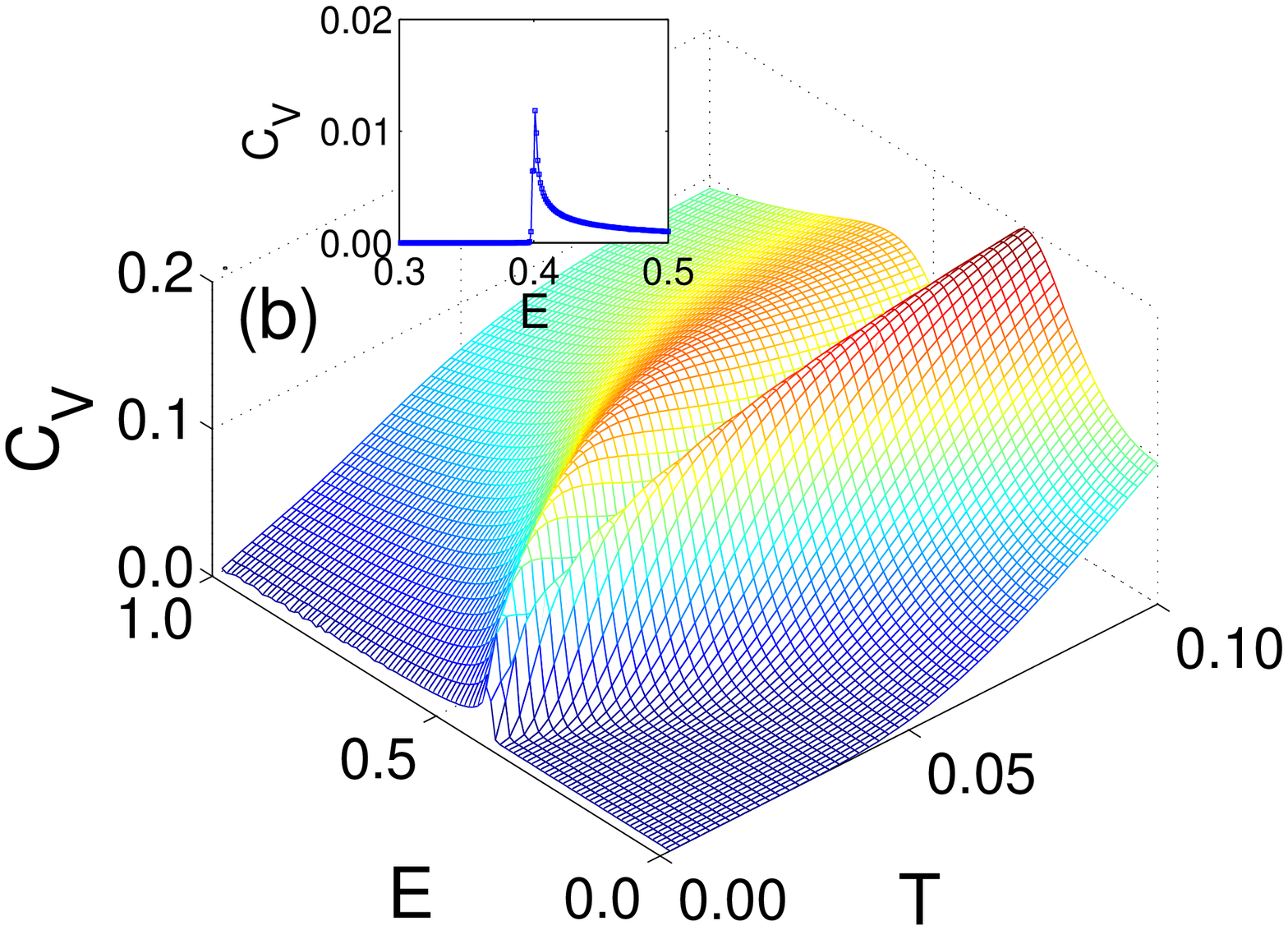}
 \caption{(Color online) The three-dimensional plot of the specific heat $C_V$ versus temperature and DM interaction for
(a) $h$=0.2, and
(b) $h$=0.8. Insets show the specific heat versus $E$ at $T=0.0005$. Parameters: $J_o$=1, $J_e$=1, $\theta=\pi/3$. }
\label{fig:cv1}
\end{figure}

\begin{figure}[t!]
\includegraphics[width=8cm]{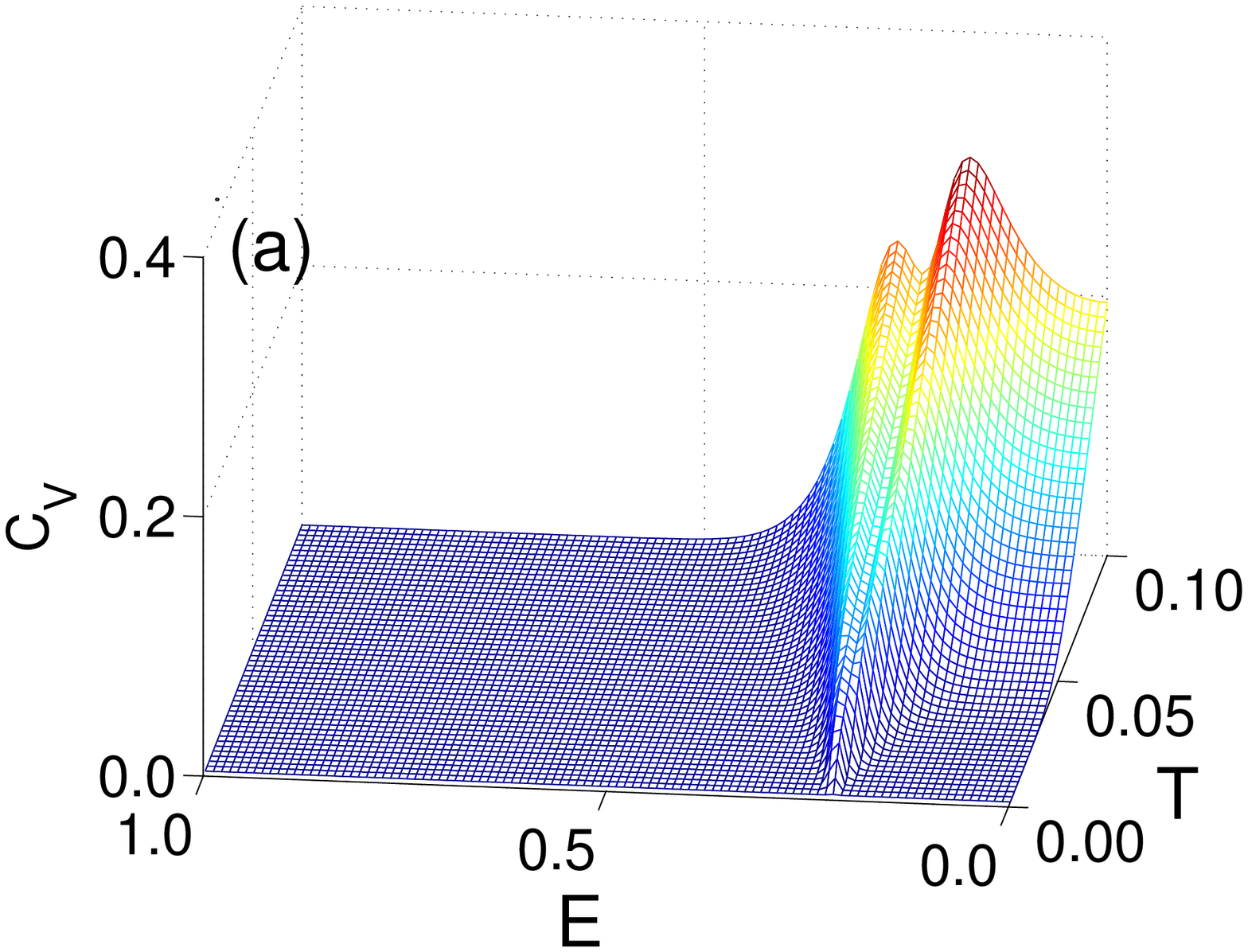}
\includegraphics[width=8cm]{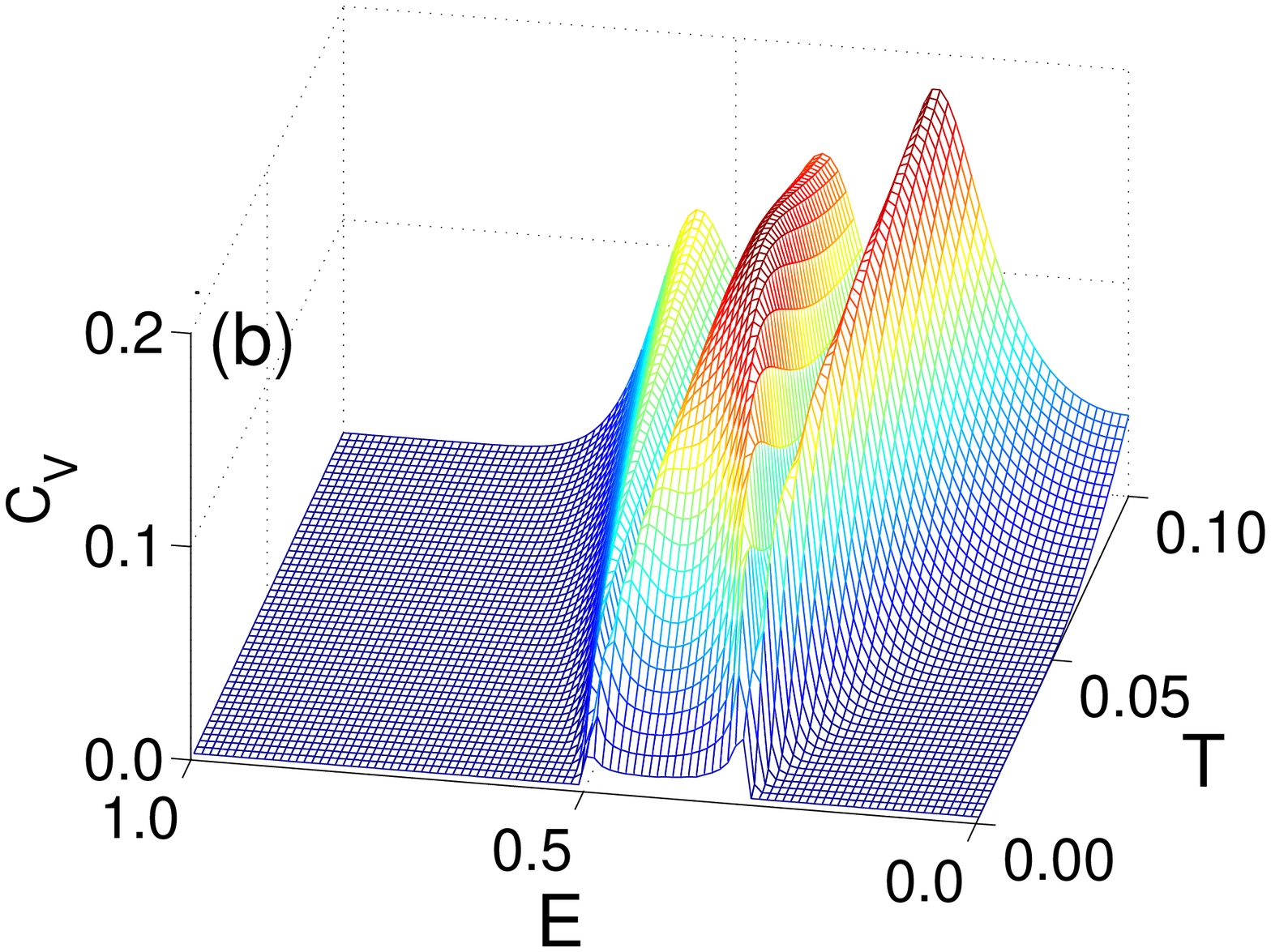}
 \caption{(Color online) The three-dimensional plot of the specific heat $C_V$ versus temperature and DM interaction for
(a) $h$=0.5, and
(b) $h$=1.5. Parameters: $J_o$=1, $J_e$=4, $\theta=\pi/3$. }
\label{fig:cv2}
\end{figure}

In addition, the low-temperature behavior of the heat capacity is readily measured, such as specific heat measurements of copper benzoate \cite{Dender97}, transverse Ising magnet CoNb$_2$O$_6$ \cite{Liang15} and
Heisenberg antiferromagneic Cu(C$_4$H$_4$N$_2$)(NO$_3$)$_2$ \cite{Kono15}.
Figure \ref{fig:cv1} shows a three-dimensional plot of the specific heat $C_V$ for $J_o$=$J_e$. The specific heat contains here a broad peak
around the QCP and develops a local minimum at the top of
the peak at the QCP for extremely low temperatures. By $C_V$=$T$ $(\partial{\cal S}/\partial T)_{E,h}$, the specific heat also follows $C_V$  $\propto T^{d/z}$ in gapless regimes, since $C_V/T $ is approximately
proportional to $D(k_B T)$. More precisely, $C_V \propto \sqrt{T}$ at critical points \cite{Kono15}and $C_V \propto T$ in gapless chiral phase\cite{Liang15}, in addition to an exponential activation in the gapped phases. Although the power-law scaling of the specific heat at the QCPs is identical to that of the entropy, the shallow trough in the specific heat implies that the critical temperature $T_c(E)$ falls to zero as $E \to E_c$. A constant-$T$ contour measured at $T=0.0005$ (see insets of Fig. \ref{fig:cv1}) converges to a prominent peak profile $C_V$ for $T=0$ at the QCP, and the relatively high temperature quenches any discernible peak feature. For $J_o$$\neq$$J_e$, two successive local minima show up in Fig.\ref{fig:cv2}, implying two QPTs with the
increase of $E$. The universal characteristic power laws in Tomonaga-Luttinger-liquid phase can be identified in the measured specific heat \cite{Kono15,Xiang98}.

\section{Conclusion}
 \label{sec:Con}
Plenteous intriguing phenomena in condensed matter systems
originate from the interplay of strong interactions and frustrations. The dominating finite-range interactions in many-body systems can lead to kaleidoscopic self-ordered phases of matter.
The one-dimensional generalized compass model mimics a distorted TM ion-oxygen ion-TM ion chain with a zigzag alignment, and
it is one of very few quantum system to be exact solvable.
In the paper, we study quantum phase transitions in this frustrated model
as tuning an external magnetic field and staggered Dzyaloshinskii-Moriya interaction arising from the distortion. We present the exact solution by means of Jordan-Wigner transformation. We study the fermionic spectra, excitation gap,  spin correlations, and  critical properties at
phase transitions.  Then we establish the phase diagram. In order to uncover quantum critical behaviors of our model near the quantum critical points,  we here  supplement a temperature dependence of a few thermodynamic quantities, including entropy and specific heat. The entropy and specific heat  exhibit  characteristic power-law behavior with low temperature $T$ due to the density of states at Fermi surface, which is determined by low-energy excitations controlled by the dynamical
exponent $z$. $z=2$ at quantum critical point while $z=1$ in  Tomonaga-Luttinger-liquid phase.

\acknowledgments
We thank G.-H. Liu for helping us plot Fig.\ref{string}. W.-L.Y. acknowledges support by the Natural Science Foundation of Jiangsu Province of China under Grant No. BK20141190 and the NSFC under Grant No. 11474211. W.-H. N. was financially supported by the NSFC under Grant No. 21473240.


\begin{references}
\bibitem{Dzyaloshinskii58} I. Dzyaloshinskii, J. Phys. Chem. Solids \textbf{4}, 241 (1958).

\bibitem{Moriya60} T. Moriya, Phys. Rev. \textbf{120}, 91 (1960).

\bibitem{Fert80}A. Fert and P. M. Levy, Phys. Rev. Lett. \textbf{44}, 1538 (1980).

\bibitem{Cheong07}S.-W. Cheong and M. Mostovoy, Nature Mater. \textbf{6}, 13 (2007).
\bibitem{Hosho05}Hosho Katsura, Naoto Nagaosa, and Alexander V. Balatsky
Phys. Rev. Lett. \textbf{95}, 057205 (2005);  M. Mostovoy, Phys. Rev. Lett. \textbf{96}, 067601 (2006).

\bibitem{Tokura10} Y. Tokura and S. Seki,
                   Adv. Mater. \textbf{22}, 1554 (2010).

\bibitem{Heinze11}S. Heinze, K. von Bergmann, M. Menzel, J. Brede, A.
Kubetzka, R. Wiesendanger, G. Bihlmayer, and S. Bl\"{u}gel,
Nat. Phys. \textbf{7}, 713 (2011).

\bibitem{Luchaire16} C. Moreau-Luchaire {\it et al.}, Nat. Nanotechnol. \textbf{11}, 444 (2016).

\bibitem{Boulle16}O. Boulle {\it et al.}, Nat. Nanotechnol. \textbf{11} , 449 (2016).

\bibitem{Jiang15} Wanjun Jiang, Pramey Upadhyaya, Wei Zhang, Guoqiang Yu, M. Benjamin Jungfleisch, Frank Y. Fradin, John E. Pearson, Yaroslav Tserkovnyak3, Kang L. Wang, Olle Heinonen, Suzanne G. E. te Velthuis, Axel Hoffmann, Science \textbf{349}, 283 (2015).

\bibitem{Bode07}M. Bode, M. Heide, K. von Bergmann, P. Ferriani, S. Heinze,
G. Bihlmayer, A. Kubetzka, O. Pietzsch, S. Bl\"{u}gel, and R.
Wiesendanger, Nature (London) \textbf{447}, 190 (2007).

\bibitem{Goff95}J.P. Goff, D.A. Tennant, S.E. Nagler, Phys. Rev. B \textbf{52},
15992 (1995).
\bibitem{Jafari08}R. Jafari, M. Kargarian, A. Langari, and M. Siahatgar
Phys. Rev. B \textbf{78}, 214414 (2008).
\bibitem{You2} W.-L. You, G.-H. Liu, P. Horsch, and A. M. Ole\'s,                    Phys. Rev. B \textbf{90}, 094413 (2014).
\bibitem{Liu15}Guang-Hua Liu, Wen-Long You, Wei Li, Gang Su, J. Phys.: Condens. Matter \textbf{27}, 165602 (2015).
\bibitem{Affleck99}I. Affleck, M. Oshikawa, Phys. Rev. B \textbf{60}, 1038 (1999).

\bibitem{Oshikawa97}Masaki Oshikawa and Ian Affleck, Phys. Rev. Lett. \textbf{79}, 2883 (1997).

\bibitem{Zhao03}J. Z. Zhao, X. Q. Wang, T. Xiang, Z. B. Su, and L. Yu
Phys. Rev. Lett. \textbf{90}, 207204 (2003).

\bibitem{Dmitrienko14}V. E. Dmitrienko, E. N. Ovchinnikova, S. P. Collins,G. Nisbet,	G. Beutier, Y. O. Kvashnin, V. V. Mazurenko, A. I. Lichtenstein and M. I. Katsnelson£¬ Nature Phys. \textbf{10}, 202 (2014).

\bibitem{Schauss15}P. Schau{\ss},
 J. Zeiher, T. Fukuhara, S. Hild, M. Cheneau, T. Macr{\`\i}, T. Pohl, I. Bloch, C. Gross, Science \textbf{347}, 1455 (2015).

\bibitem{You1} W.-L. You, P. Horsch, and A. M. Ole\'s, Phys. Rev. B \textbf{89}, 104425 (2014).

\bibitem{You16}W.-L. You, Y.-C. Qiu, and A. M. Ole\'s, Phys. Rev. B \textbf{93}, 214417 (2016).

\bibitem{Mochizuki11} Masahito Mochizuki, Nobuo Furukawa, and Naoto Nagaosa, Phys. Rev. Lett. \textbf{105}, 037205 (2010); Phys. Rev. B \textbf{84}, 144409 (2011).

\bibitem{Qiu16} Y.-C. Qiu, Q.-Q. Wu and W.-L. You, J. Phys.: Condens. Matter, \textbf{28}, 496001(2016).
\bibitem{Schotte98} U. Schotte, A. Kelnberger, N. Stsser, J. Phys.: Condens. Matter \textbf{10}, 6391 (1998).

\bibitem{EBarouch70} E. Barouch and B. M. McCoy,
                   Phys. Rev. A  \textbf{2}, 1075 (1970);
                                 \textbf{3}, 786 (1971).
\bibitem{Tong15}Y. Xiong and P. Tong, New J. Phys. \textbf{17}, 013017 (2015);X. Wang, T. Liu, and Y. Xiong and P. Tong, Phys. Rev. A \textbf{92}, 012116 (2015).

\bibitem{Jafari17}R. Jafari and Henrik Johannesson, Phys. Rev. Lett. \textbf{118}, 015701 (2017).

\bibitem{Altland} A. Altland and M. R. Zirnbauer,
                   Phys. Rev. B \textbf{55}, 1142 (1997).

\bibitem{Chiu16}Ching-Kai Chiu, Jeffrey C.Y. Teo, Andreas P. Schnyder, and Shinsei Ryu, Rev. Mod. Phys. \textbf{88}, 035005 (2016).

\bibitem{Kitaev01}A. Yu. Kitaev, Phys.-Usp. (Suppl.)  \textbf{44}, 131 (2001).

\bibitem{Ghosh10}P. Ghosh, J. D. Sau, S. Tewari, and S. Das Sarma, Phys. Rev. B
\textbf{82}, 184525 (2010).

\bibitem{Sergienko06} I. A. Sergienko and E. Dagotto,
                   Phys. Rev. B  \textbf{73}, 094434 (2006).

\bibitem{Vidal07} G. Vidal,
                   Phys. Rev. Lett. \textbf{98}, 070201 (2007); R. Or\'us   and G. Vidal, Phys. Rev. B  \textbf{78}, 155117 (2008).

\bibitem{Liu13} G.-H. Liu, Wei Li, W.-L. You, G. Su, and G.-S. Tian,
                   Eur. Phys. J. B  \textbf{86}, 227 (2013); G.-H. Liu, Wei Li, W.-L. You, G.-S. Tian, and  G. Su, Phys. Rev. B  \textbf{85}, 184422 (2012).


\bibitem{Rost09} A. W. Rost, R. S. Perry, J.-F. Mercure, A. P. Mackenzie,
                   and S. A. Grigera,
                   Science \textbf{325}, 1360 (2009).
 \bibitem{Dender97} D. C. Dender, P. R. Hammar, Daniel H. Reich, C. Broholm, and G. Aeppli, Phys. Rev. Lett. \textbf{79}, 1750 (1997).

\bibitem{Liang15}Tian Liang, S. M. Koohpayeh, J. W. Krizan, T. M. McQueen, R. J.  Cava and N. P. Ong, Nature Commun. \textbf{6}, 7611 (2015).

 \bibitem{Kono15}Y. Kono, T. Sakakibara, C.P. Aoyama, C. Hotta, M.M. Turnbull, C.P. Landee, and Y. Takano, Phys. Rev. Lett. \textbf{114}, 037202 (2015).

\bibitem{Xiang98}T. Xiang, Phys. Rev. B \textbf{58}, 9142 (1998).


%
\end{references}
\end{document}